\input{aipcheck}

\documentclass[
    ,final            
  ]
  {aipproc}

\layoutstyle{6x9}

\begin{document}

\title{Hunting for asymptotia at LHC}



\author{G. Pancheri}{
  address={INFN Frascati National Laboratories, Via E. Fermi 40, 00444, Italy},
  email={pancheri@lnf.infn.it}
}

\author{D.A. Fagundes}{address={INFN Frascati National Laboratories, Via E. Fermi 40, 00444, Italy},
altaddress={Instituto de F\'{\i}sica Gleb Wataghin, Universidade Estadual de
Campinas, UNICAMP, 13083-859 Campinas SP, Brazil},email={fagundes@ifi.unicamp.br}}

\author{A. Grau}{
  address={Departamento de F\'\i sica Te\'orica y del Cosmos, Universidad de
Granada, 18071 Granada, Spain},
   email={igrau@ugr.es},
}
\author{S. Pacetti} {
  address={Physics Department and INFN, University of Perugia, 06123 Perugia, Italy },
  email={simone.pacetti@pg.infn.it},
}
\author{Y.N. Srivastava}{
address={Physics Department and INFN, University of Perugia, 06123 Perugia, Italy },
  email={yogendra.srivastava@pg.infn.it}}

\begin{abstract}
We discuss whether the  behaviour of some hadronic quantities, 
such as the total cross-section, the ratio of the elastic to the total 
cross-section, are presently exhibiting the asymptotic behaviour 
expected at very large energies. We find phenomenological evidence 
that at LHC7 there is still space for further evolution.
\end{abstract}

\maketitle


\section{Introduction}
The TOTEM measurement of the total and elastic differential cross-section at 
LHC~\cite{Antchev:2012prepelastic}
has challenged some models, but also confirmed older predictions. The question addressed in this contribution is  whether we have reached the region where asymptotic theorems are valid, in particular whether the Froissart bound is saturated with   $\sigma_{total}$  rising as $(\ln s)^2$, and whether the elastic cross-section is closer to the black disk limit, for which $R_{el}=\sigma_{elastic}/\sigma_{total}\rightarrow 1/2$.  In our opinion, it is too early to claim that we have reached asymptotia, and in this contribution will discuss our findings.
\section{The total cross-section: fast rise and saturation}
While it is evident that to describe the energy dependence of  the total cross-section  it is necessary to study the scattering dynamics in the large impact parameter limit, a direct link between confinement and the total cross-section limiting behaviour is yet to be obtained from first principles. A  model of soft gluon resummation, developed through a number of years and summarized here,  represents an effort in this direction.
\subsection{The fast rise and the  mini-jet contribution }
The model we shall describe in this section,  belongs to the general category of mini-jet models, which attribute the rise of the total cross-section to the onset of perturbative QCD (pQCD) effects and estimate the rise from  the  cross-section for parton-parton scattering at LO between two hadrons. This can be accomplished using library available  Parton Density Functions (PDFs), with standard DGLAP evolution. These mini-jet cross-sections rise very rapidly, like $s^\epsilon$ with $\epsilon \sim 0.3-0.4$, posing a problem. The counterpart of the  mini-jets rise 
is  the hard Pomeron, behaving as $s^\Delta$ in Reggeon field theory models. This rise has a similar origin in the two approaches, namely the increasing contribution of (low momentum)  low-$x $ partons to hard and semi-hard scattering. The contribution of parton-parton scattering to the rise of the total cross-section was advocated as soon as ISR experiments confirmed the rise with energy of $\sigma_{total}$~\cite{Cline:1973kv}. The fact that semi-hard  interactions played an important role in the total cross-section and inelastic interactions was  amply discussed in the mid '80~\cite{Gribov:1984tu,Pancheri:1985rm} in pQCD context.

 Our approach differs with respect to Reggeon  models since  phenomenologically determined PDFs (at LO) are used for the hard scattering, thus keeping connection to standard pQCD. For scattering of hadrons A and B, we calculate the mini-jet cross-sections as
\begin{equation}
 \sigma^{AB}_{\rm jet} (s,p_{tmin})=
\int\limits_{p_{tmin}}^{\sqrt{s}/2} d p_t \hspace{-0.2cm}\int
\limits_{4 p_t^2/s}^1 d x_1 \hspace{-0.2cm}
\int\limits_{4 p_t^2/(x_1 s)}^1 d x_2 \sum_{i,j,k,l} f_{i|A}(x_1,p_t^2)
f_{j|B}(x_2,p_t^2)
  \frac { d \hat{\sigma}_{ij}^{ kl}(\hat{s})} {d p_t}
\label{eq:minijet}
\end{equation}
In Eq.~(\ref{eq:minijet}), $p_{tmin}$ is a fixed scale introduced to separate perturbative from non-perturbative processes, namely only scattering with   $p_t>p_{tmin}$ is described by the mini-jet cross-section.  The PDFs are DGLAP evoluted with scale $p_t^2$ and the parton-parton cross-sections are calculated with the asymptotic freedom expression for $\alpha_s(p_t^2)$.
 
To ensure unitarity, we then formulate the problem within the eikonal representation.  The minijet cross-section is used as input to the average number of collisions at impact parameter $\bf b$ and subsequently  to the eikonal function,  which will give the scattering amplitude at $t=0$, i.e.
\begin{eqnarray}
\sigma_{total}&=&2\int d^2{\bf b}[1-e^{-\bar n(s,b)/2}]\\
\bar n(s,b)&=& n_{soft}(s,b)+A_{hard}(b,s)\sigma_{jet}(s,p_{tmin})
\label{eq:nbar}
\end{eqnarray}
In Eq.~(\ref{eq:nbar}) $n_{soft}$ is the average number of collisions contributing to the total cross-section at low cm energies such as  $\sqrt{s}\sim  5-10$ GeV, when the mini-jet contribution is very small. $A_{hard}(b,s)$ is the impact parameter distribution for partons averaged over the parton densities \cite{Grau:1999em}.
\subsection{Gluon saturation from soft $k_t$ resummation}
The impact parameter dependence in eikonal models plays the major role in softening the rise of the total cross-section, with the infrared behaviour of soft gluons emitted in the process leading  to saturation. In Reggeon type models this is accomplished through soft Pomeron contributions, with the Pomeron trajectory arising through exponentiation of soft contributions. 
The major difference between the approach developed in Refs.~\cite{Grau:1999em,Godbole:2004kx} and Reggeon models lies in the inclusion of the infrared region in resummation. To deal with this region, an effective quark-gluon coupling has been proposed as
\begin{equation}
\alpha_{eff}(k_t^2)\rightarrow \left(\frac{\Lambda^2}{k_t^2}\right)^p\,, \ \ \ \ \ \ \ \ k_t<<\Lambda \label{eq:alphasing}
\end{equation}
where $\Lambda $ is the QCD scale and the parameter $p$, which regulates the infrared divergence, is restricted to the range  $1/2<p<1$. 
To cover the full range of values accessed in resummation, an interpolating 
function between the asymptotic freedom expression and the infrared behaviour 
can be used~\cite{Corsetti:1996wg}.

The  acollinearity effect produced by soft gluon emission, accompanying  the mini-jet cross-sections,  reduces the probability of collision, as partons do not see each other head-on anymore. Thus soft gluon emission can lead  to saturation. To include it in the model for the total cross-section, we make the ansatz that the impact parameter distribution in  hard collisions is obtained from   the Fourier transform of resummed soft gluon emission, 
\begin{equation}
A_{hard}(b,s)\equiv A_0(s,p_{tmin}) e^{
-h(b,s,p_{tmin})
}=\frac{
e^{
-h(b,s,p_{tmin})
}
}
{\int d^2 {\bf b}\  e^{
-h(b,s,p_{tmin})
}
}
\end{equation}
with 
\begin{equation}
h(b,s,p_{tmin})=\frac{8}{3\pi^2}\int_0^{q_{max}} d^2{\bf k}_t \frac{\alpha_s(k_t^2)}{k_t^2}\ln \frac{2 q_{max}(s,p_{tmin})}{k_t}[1-J_0(k_t b)]
\end{equation}
When the singular, but integrable, coupling of Eq.~(\ref{eq:alphasing})  is used in the single soft gluon function $h(b,s,p_{tmin})$, the saturation effect, briefly described here,   changes the fast rise due to hard processes,  given by $s^\epsilon $, into a slower rise, as we show next. 

We notice a recent attempt by 
Fagundes-Luna-Menon-Natale~\cite{Fagundes:2011zx}, in  the
\textit{Dynamical Gluon Mass} approach,  to include an infrared  frozen coupling constant  in  resummation models. 
\subsection{Onset of the Froissart limit}
\begin{figure}[h!]
\includegraphics[width=72mm,height=66mm]{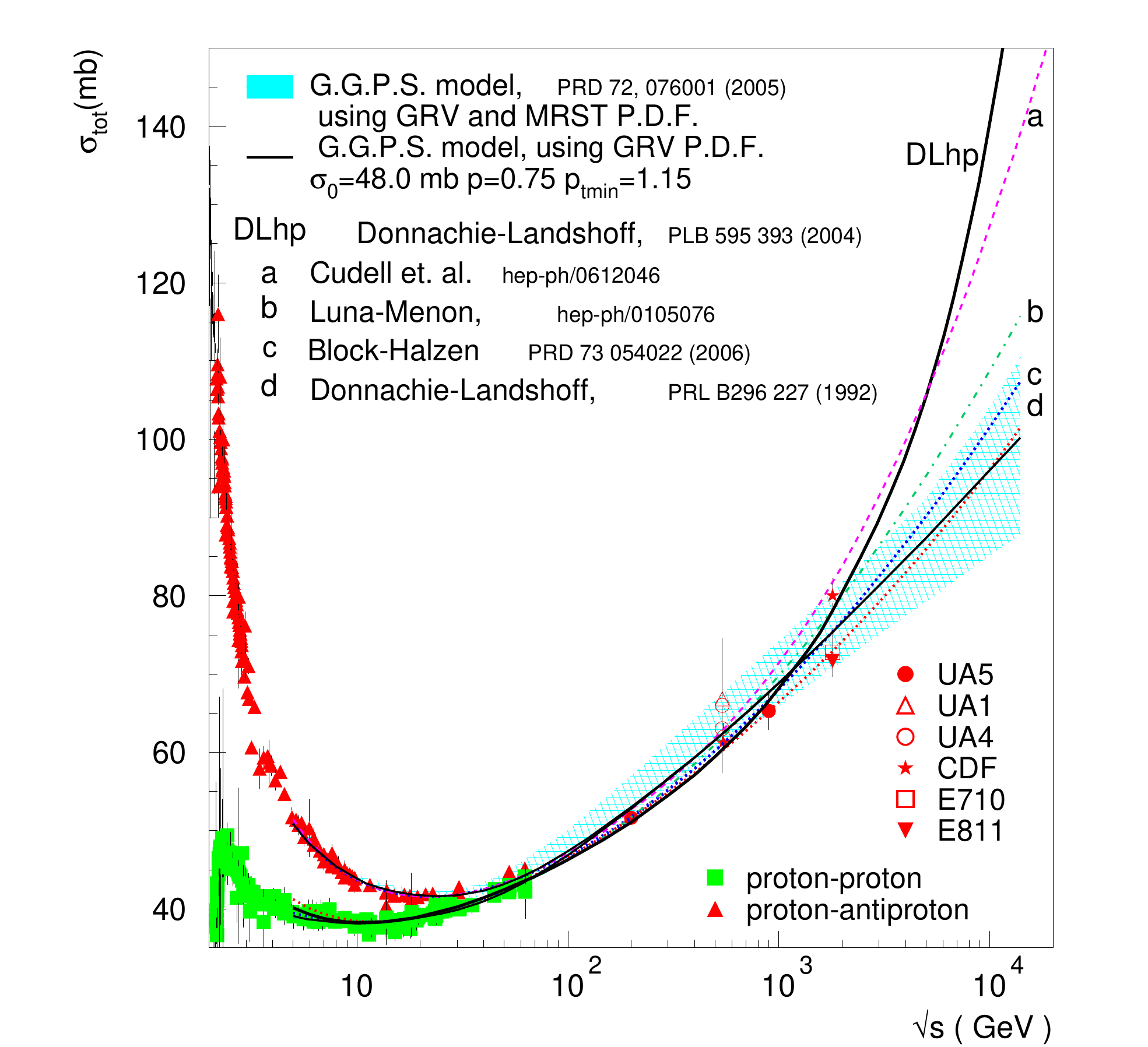}
\hspace{5mm}
\includegraphics[width=70mm,height=70mm]{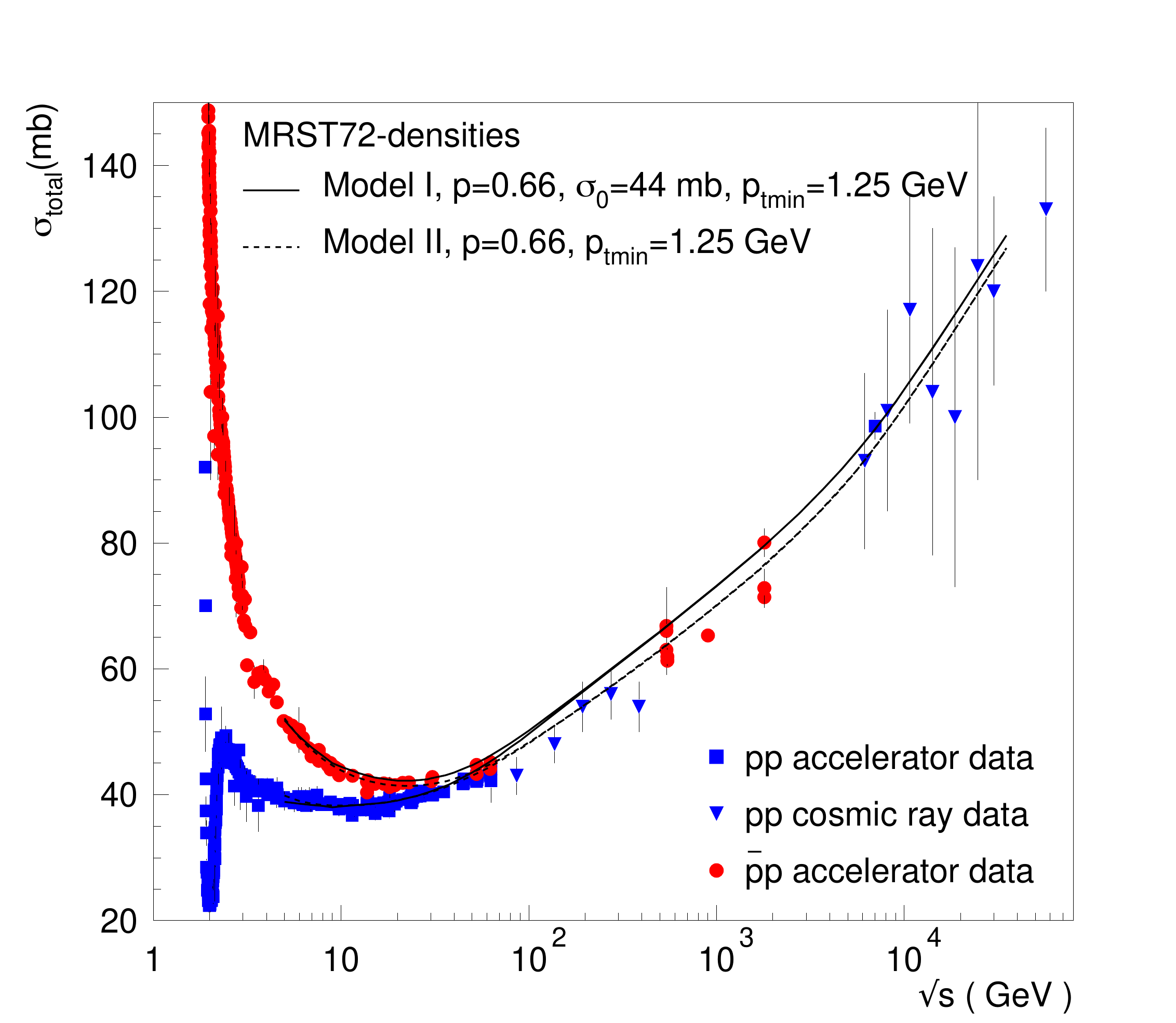}
  \caption{Data for total $pp$ and ${\bar p}p$ cross-section compared with expectations from the 
  eikonal mini-jet model with  soft $k_t$-gluon resummation described in the text. At left, the blue band gives the  predictions of this model from \cite{Achilli:2007pn}, at right we show a  compilation of most recent data from accelerator and cosmic ray experiments~\cite{Beringer:1900zz,Collaboration:2012wt} compared with LO parametrization of the model with MRST72 densities. The two curves correspond to different parametrizations for $n_{soft}$.}
\label{fig:sigtot}
\end{figure}

The softening effect of resummation can be seen to arise through the fact that resummation in the infrared region induces   a strong cut-off in impact parameter space, i.e.
\begin{equation}
A_{hard}(b,s)\sim e^{
-(b{\bar \Lambda)}^{2p}
}
\ \ \ \ \ b\rightarrow \infty
\end{equation}
and  the high energy limit of the total cross-section becomes \cite{Grau:2009qx}
\begin{equation}
  \sigma _T (s) \approx 2\pi \int_0^\infty  {db^2 } [1 - e^{ -
C(s)e^{ - (b{\bar \Lambda})^{2p} } } ] \rightarrow (\epsilon \ln s )^{1/p}
\label{eq:sigT}
\end{equation} 
 where $2C(s) \propto A_0(s)  (s/s_0 )^\varepsilon$. Phenomenologically, we see that, upon eikonalization and parametrization of the soft contribution  $n_{soft}$, a good description of the behavior of the total cross-section appears.  Fig.~\ref{fig:sigtot} shows the results of this model, with, at left,   predictions from this and other models  before LHC~\cite{Achilli:2007pn}, and, at right,  the result of this model for the particular choice of MRST72 PDFs (upper curve of blue band), compared with LHC 
recent data  and cosmic ray results. 

Our phenomenology indicates that for the LO choice of MRST72 densities, the value $p=2/3$ gives a good description of data up to the highest value available. This is not a fit to the data, but the application of a specific model. However,  this result may indicate that there is still space for the   saturation of the Froissart bound. 
\begin{figure}[h!]
\includegraphics[width=120mm]{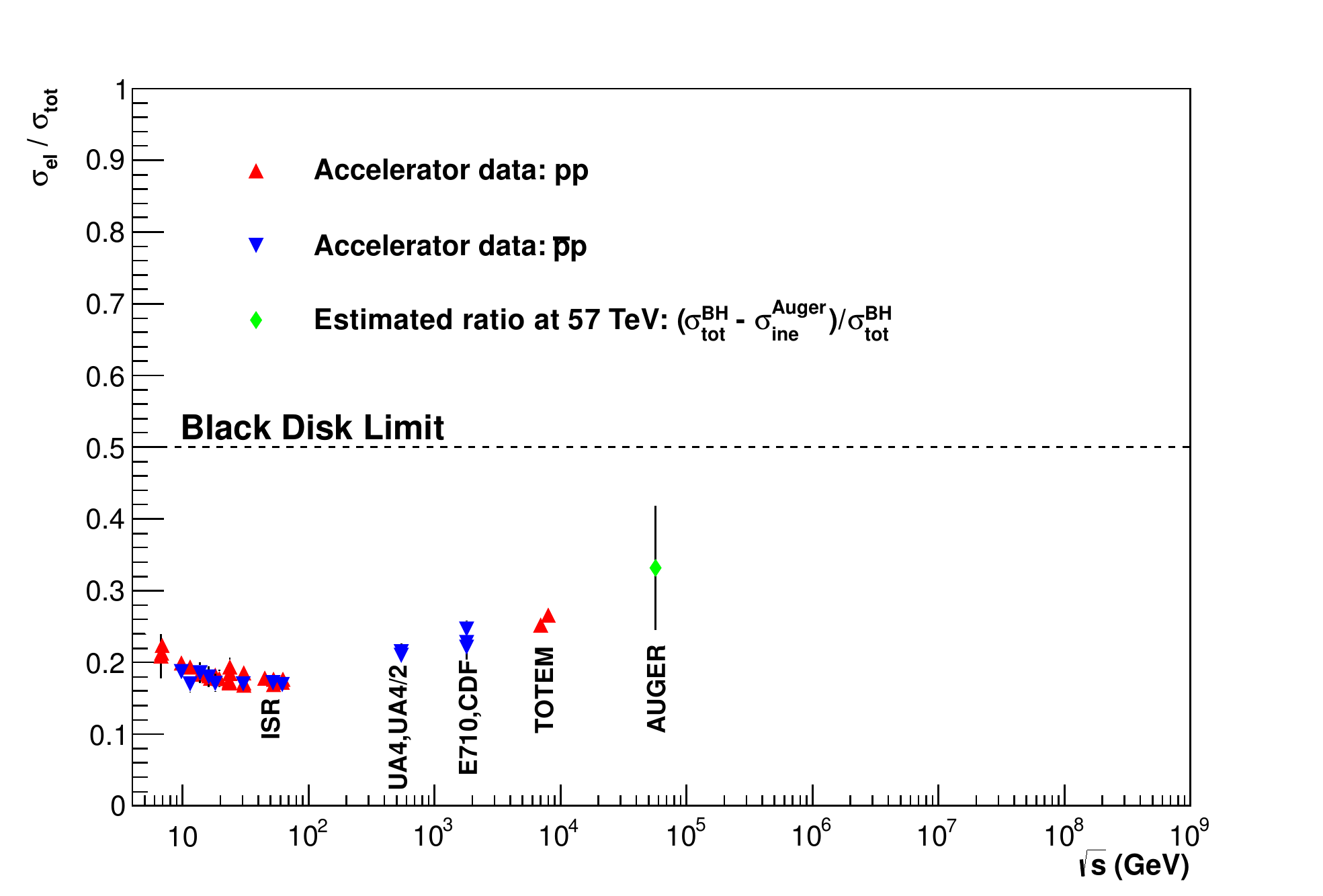}
  \caption{The ratio $\sigma_{elastic}/\sigma_{total}$ from accelerator data \cite{Beringer:1900zz} up to  AUGER measurement \cite{Collaboration:2012wt}.}
  \label{fig:rel}
  \end{figure}
\section{The elastic  cross-section and the black disk limit}
Here we turn to the elastic cross-section and in Fig.~\ref{fig:rel} show our compilation for $R_{el}=\sigma_{elastic}/\sigma_{total}$. This figure includes the latest TOTEM values~\cite{Antchev:2012prepelastic,Antchev:2012prep8TeV} as well as the result from the AUGER collaboration for the inelastic cross-section $\sigma_{inelastic}^{AU}$~\cite{Collaboration:2012wt}. For the latter, we have used the value for $\sigma_{total}$ obtained  by Block and Halzen $\sigma_{total}^{BH}(57\ {\rm TeV})=(134.8 \pm 1.5 )\ {\rm mb}$~\cite{Block:2011vz} and thus estimated the ratio $R_{elastic}=(\sigma_{total}^{BH}-\sigma_{inelastic}^{AU})/\sigma_{total}^{BH}$ at 57 TeV. The figure clearly indicates that the black disk limit has not been reached yet. Notice that inclusion of diffraction changes this limit, as discussed in Ref.~\cite{Grau:2012wy}. For a model independent fit to  the data,  see Ref.~\cite{Fagundes:2011hv}.
\begin{theacknowledgments}
AG acknowledges partial support by Spanish MEC (FPA2010-16696,
AIC-D-2011-0818) 
 and by Junta de Andalucia (FQM 03048, FQM 6552, FQM 101). DAF thanks the Brazilian Funding Agency FAPESP for financial support
(contract: 2012/12908-4).
\end{theacknowledgments}



\bibliographystyle{aipproc}
\bibliography{ref-pancheri1,diff,ref_exps}

\end{document}